\documentclass[prl,twocolumn,showpacs,preprintnumbers,amsmath,amssymb]{revtex4}
\usepackage{dcolumn}
\usepackage{multirow}
\usepackage{epsfig}
\usepackage[english]{babel}

\begin{document}

\bibliographystyle{try}
\topmargin 0.1cm

\newcounter{univ_counter}
\setcounter{univ_counter} {0}
\addtocounter{univ_counter} {1}
\edef\HISKP{$^{\arabic{univ_counter}}$ } \addtocounter{univ_counter}{1}
\edef\GATCHINA{$^{\arabic{univ_counter}}$ } \addtocounter{univ_counter}{1}
\edef\ERLANGEN{$^{\arabic{univ_counter}}$ } \addtocounter{univ_counter}{1}
\edef\PI{$^{\arabic{univ_counter}}$ } \addtocounter{univ_counter}{1}
\edef\KVI{$^{\arabic{univ_counter}}$ } \addtocounter{univ_counter}{1}
\edef\BOCHUM{$^{\arabic{univ_counter}}$ } \addtocounter{univ_counter}{1}
\edef\DRESDEN{$^{\arabic{univ_counter}}$ } \addtocounter{univ_counter}{1}
\edef\BASEL{$^{\arabic{univ_counter}}$ } \addtocounter{univ_counter}{1}
\edef\GIESSEN{$^{\arabic{univ_counter}}$ } \addtocounter{univ_counter}{1}

\title{Neutral pion photoproduction off protons in the energy range 0.3\,GeV$\,<\,{\mathbf E}_{\boldsymbol\gamma}\,<$\,3\,GeV\\
\vskip 2mm
}
\author{O.~Bartholomy~\HISKP,
V. Cred\'e~\HISKP$^{\diamond}$,
H.~van~Pee~\HISKP$^*$,
A.\,V.~Anisovich~$^{1,2}$,
G.~Anton~\ERLANGEN,
R.~Bantes~\PI,
Yu.~Beloglazov~\GATCHINA,
R.~Bogend\"orfer~\ERLANGEN,
R.~Castelijns~\KVI,
A.~Ehmanns~\mbox{\HISKP,}
J.~Ernst~\HISKP,
I. Fabry~\HISKP,
H.~Flemming~\BOCHUM,
A.~F\"osel~\ERLANGEN,
H.~Freiesleben~\mbox{\DRESDEN,}
M.~Fuchs~\HISKP,
Ch.~Funke~\HISKP,
R.~Gothe~\PI$^+$,
A.~Gridnev~\GATCHINA,
E.~Gutz~\HISKP,
S.\,K.~H\"offgen~\PI,
I.~Horn~\HISKP,
J.~H\"o\ss l~\mbox{\ERLANGEN,}
R.~Joosten~\HISKP~,
J.~Junkersfeld~\HISKP~,
H.~Kalinowsky~\HISKP,
F.~Klein~\PI,
E.~Klempt~\HISKP,
H.~Koch\mbox{~\BOCHUM,}
M.~Konrad~\mbox{\PI,}
B.~Kopf~\mbox{\BOCHUM\DRESDEN,}
B.~Krusche~\BASEL,
J.~Langheinrich~\PI$^+$,
H.~L\"ohner~\KVI,
I.~Lopatin~\mbox{\GATCHINA,}
J.~Lotz~\HISKP,
H.~Matth\"ay~\BOCHUM,
D.~Menze~\mbox{\PI,}
J.~Messchendorp~\GIESSEN$^\dagger$,
C.~Morales~\PI,
D.~Novinski~\GATCHINA,
M.~Ostrick~\mbox{\PI,}
A.~Radkov~\mbox{\GATCHINA,}
J.~Reinnarth~\HISKP,
A.\,V.~Sarantsev~$^{1,2}$,
S.~Schadmand~\GIESSEN,
Ch.~Schmidt~\HISKP,
H.~Schmieden~\PI,
B.~Schoch~\PI,
G.~Suft~\ERLANGEN,
V.~Sumachev~\GATCHINA,
T.~Szczepanek~\HISKP,
U.~Thoma~\HISKP$^*$,
D.~Walther~\PI and
Ch.~Weinheimer~\HISKP \\
(The CB--ELSA Collaboration)}
\affiliation{$^1$ Helmholtz--Institut f\"ur Strahlen-- und Kernphysik,
Universit\"at Bonn, Germany}
\affiliation{\GATCHINA Petersburg
Nuclear Physics Institute, Gatchina, Russia}
\affiliation{\ERLANGEN Physikalisches Institut,
Universit\"at Erlangen, Germany}\affiliation{\PI Physikalisches Institut,
Universit\"at Bonn, Germany}
\affiliation{\KVI Kernfysisch Versneller Instituut, Groningen, Netherlands}
\affiliation{\BOCHUM
Institut f\"ur Experimentalphysik I, Ruhr--Universit\"at Bochum, Germany}\affiliation{\DRESDEN
Institut f\"ur Kern-- und Teilchenphysik, Universit\"at
Dresden, Germany}\affiliation{\BASEL Physikalisches Institut, Universit\"at
Basel, Switzerland}
\affiliation{\GIESSEN Physikalisches Institut, Universit\"at
Gie{\ss}en, Germany\\
$^{\diamond}$ currently at Cornell University, USA, $^*$ currently at \GIESSEN, $^\dagger$ currently at \KVI, $^+$ currently at University of South Carolina,USA}

\date{\today}

\begin{abstract}
Single $\pi^0$ photoproduction has been studied with the
CB--ELSA experiment at Bonn using tagged photon energies between 0.3 and
3.0\,GeV.  The experimental setup covers a very large solid angle of
$\sim$98\% of $4\pi$. Differential cross sections
$\rm {d\sigma}/{d\Omega}$ have been measured.  Complicated
structures in the angular distributions indicate a variety of
different resonances being produced in the $s$ channel intermediate
state $\rm\gamma p\rightarrow N^{\star}(\Delta^{\star})\rightarrow
p\pi^0$.  A combined analysis including the data
presented in this letter along with other data sets reveals  contributions
from known resonances and evidence for a new resonance
$\rm N(2070)D_{15}$.
\end{abstract}
\pacs{14.20}
\maketitle


The study of baryon resonances, of their masses, widths, and decay
modes, is a cornerstone to improve our understanding of the internal
structure of nucleons. Apart from the recently suggested \mbox{exotic}
pentaquark state $\Theta^+(1540)$~\cite{Klempt:2004yz}, the pattern of
known baryon resonances below masses of 1.8\,GeV
listed by the {\bf P}article {\bf D}ata {\bf G}roup (PDG)~\cite{Hagiwara:fs} is generally well reproduced by present quark model 
calculations~\cite{Capstick:1993kb,Glozman:1997ag,Loring:2001kx}.  Above this mass,
models predict many more resonances than have been seen
experimentally. Moreover, the mass predictions of models using
one--gluon exchange~\cite{Capstick:1993kb} or instanton--induced
interactions~\cite{Loring:2001kx} no longer agree.  Lattice gauge
calculations reproduce the masses
of ground--state baryons rather well~\cite{Young:rx} and even first excited states have been simulated~\cite{Richards:2002yh}. However the structure of baryons
is still far from being understood.
\par
Results of elastic and inelastic $\rm\pi N$ scattering experiments
provide the largest fraction of our knowledge on the excitation
spectra of the nucleon and the $\rm\Delta(1232)P_{33}$.
Most excited states are also reached
in photonuclear reactions, yielding
information on their photocouplings.
The investigation of the neutral pion channel is particularly
interesting since background terms from direct production off the pion
cloud or from  production via $t$--channel exchange of pions are
suppressed.
\par
Experimental information on $\pi^0$ photoproduction is
sparse at high photon energies.  Our existing knowledge is
condensed into the parameterizations of SAID, a partial wave analysis
program run at GWU~\cite{SAID,Arndt:2003} and of MAID, the Mainz
unitarity isobar model for pion photo-- and electroproduction~\cite{MAID}.  Most data sets cover the low--energy region or date back
to the 1970s and are limited in solid angle coverage
and accuracy.  Here, we just
quote the recent JLab data on single $\pi^0$ photoproduction, covering
the photon energy range from 434 to 1742\,MeV and an angular range
which varies from $-0.89$ to 0.15 in $\cos\theta_{\rm cm}$ at the
lowest, and from $-0.79$ to 0.89 at the highest energy~\cite{Heimberg:2001}.
\par
\begin{figure}
\epsfig{file=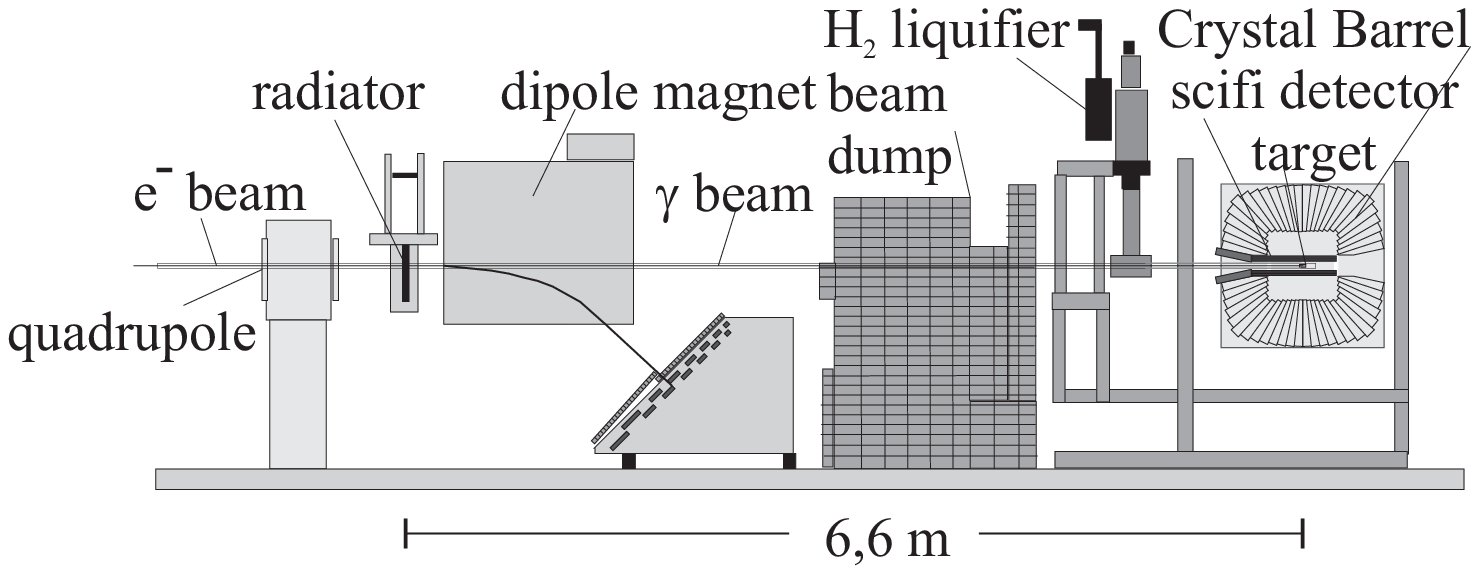,width=0.42\textwidth,clip=}
\hspace*{5.8cm}\begin{minipage}[c]{2.6cm}
\vspace*{-20mm}~\\
\caption{Experimental setup at ELSA in Bonn}
\label{cbelsa}
\vspace*{-30mm}
\end{minipage}
\begin{minipage}[c]{5.8cm}
\hspace*{-9mm}\epsfig{file=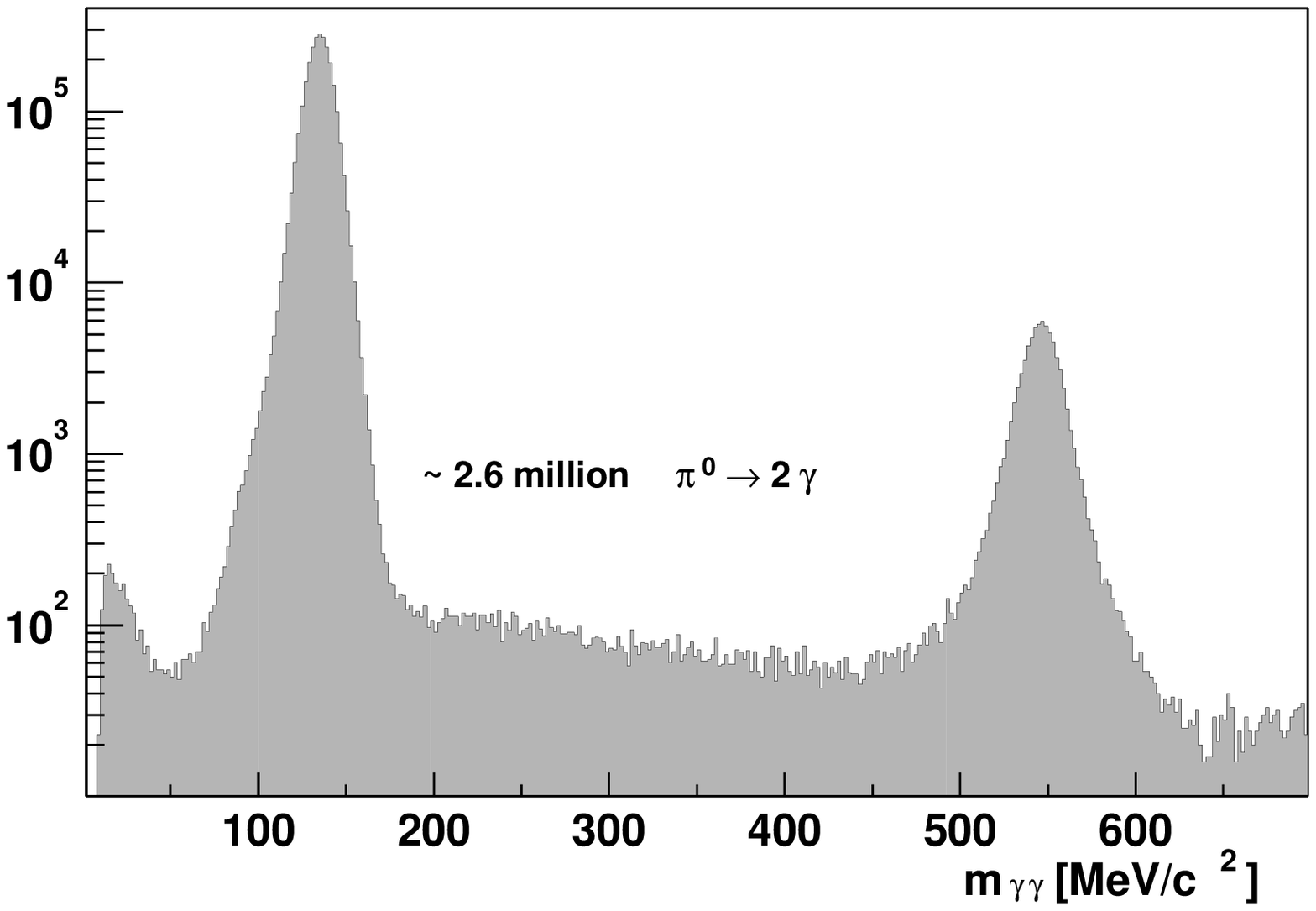,width=5.6cm}
\end{minipage}
\begin{minipage}[c]{2.7cm}
\vspace*{1.cm}
\caption{\label{figure_mass} Spectrum of invariant $\gamma\gamma$ masses on a logarithmic scale.}
\end{minipage}
\vspace*{-7mm}
\end{figure}

In this letter we report on a new measurement of single $\pi^0$
photoproduction off protons
covering the energy range from the $\rm\Delta(1232)P_{33}$ mass region
up to an invariant mass of 2.54 GeV/$c^2$.
The experiment (Fig.~\ref{cbelsa}) was carried out at the tagged
photon beam of the {\bf EL}ectron {\bf S}tretcher {\bf A}ccelerator
(ELSA) at the University of Bonn. Electrons were extracted at energies
of 1.4 and 3.2\,GeV, covering photon energies from 0.3 to
3.0\,GeV, with a typical intensity of 1--3$\,\times
10^6$\,tagged photons/s. The photon beam hit a liquid H$_2$ target of 5\,cm
length and 3\,cm diameter. Charged reaction products were detected in
a three--layer scintillating fiber (scifi) detector. One of the layers was
straight, the fibers of the other two layers encircled the target with
$\pm 25^{\circ}$ with respect to the first layer. The
intersection point of a proton could be reconstructed for polar angles
from $15^{\circ}$ to $165^{\circ}$. Charged particles and photons were
detected in the Crystal Barrel detector~\cite{Aker:1992ny}. It
consisted of 1380 CsI(Tl) crystals with photodiode readout covering
98\% of $4\pi$.  A total absorption photon detector (oil $\rm
\check{C}$erenkov counter, not shown in Fig. 1)
placed downstream was used to determine the photon flux.
\par
The coincidence between tagger and scifi detector provided the
first--level trigger of the experiment. From the hit pattern in the
Crystal Barrel detector, a fast cluster logic determined the number of
'particles' defined by clusters of contiguous crystals with individual
energy deposits above 15\,MeV.  A second level trigger was generated
for events with two or more 'particles' in the cluster logic.
In the data analysis, clusters with two local maxima were split into
two 'particles' sharing the total energy deposit.  The offline
threshold for accepted particles was set to 20\,MeV.
The proton kinetic energy had to exceed
35\,MeV to traverse the inner two scifi layers and to
produce a trigger. A proton energy of 90\,MeV was
needed to reach the barrel calorimeter and to deposit the minimum
cluster energy of 20\,MeV.
\par
In the first step of data reduction, we required both photons
to be detected in the Crystal Barrel and that not more than one
candidate for the proton was found in barrel and/or inner detector.
Thus, we retained events with two
or three 'particles'. Proton candidates were identified by the
geometrical relation of impact points in the scifi and in the
barrel. In the further analysis, protons were not explicitely
used but treated as
missing particles. The remaining two 'particles' were treated as
photons. These events were subjected to kinematic fits imposing
energy and momentum conservation by fitting
the $\rm p(\gamma,\gamma\gamma)p_{missing}$ hypothesis using
one constraint (1\,C) and the $\rm p(\gamma,\pi^0)p_{missing}$
hypothesis (2\,C) for $E_{\gamma}<$1.3\,GeV.
\par
Fig.~2 shows the $\gamma\gamma$ invariant mass spectrum after a
$10^{-4}$ confidence level cut in the 1\,C kinematic
fit. The $\pi^0$ and $\eta$ signals are observed above a small residual
background.  There are $2.6\times 10^6$ events due to $\rm\gamma
p\rightarrow p\pi^0$. The data is listed numerically in the Durham data
base~\cite{durham}.  The photoproduction of $\eta$ mesons
is discussed in a subsequent letter~\cite{Crede:2003ax}.

\par
The detector acceptance was determined from GEANT--based Monte
Carlo simulations. It
vanishes for forward protons leaving the Crystal Barrel through the
forward hole, and for protons with very low lab momenta.
\par
\begin{figure*}
\begin{minipage}{0.95\textwidth}
\includegraphics{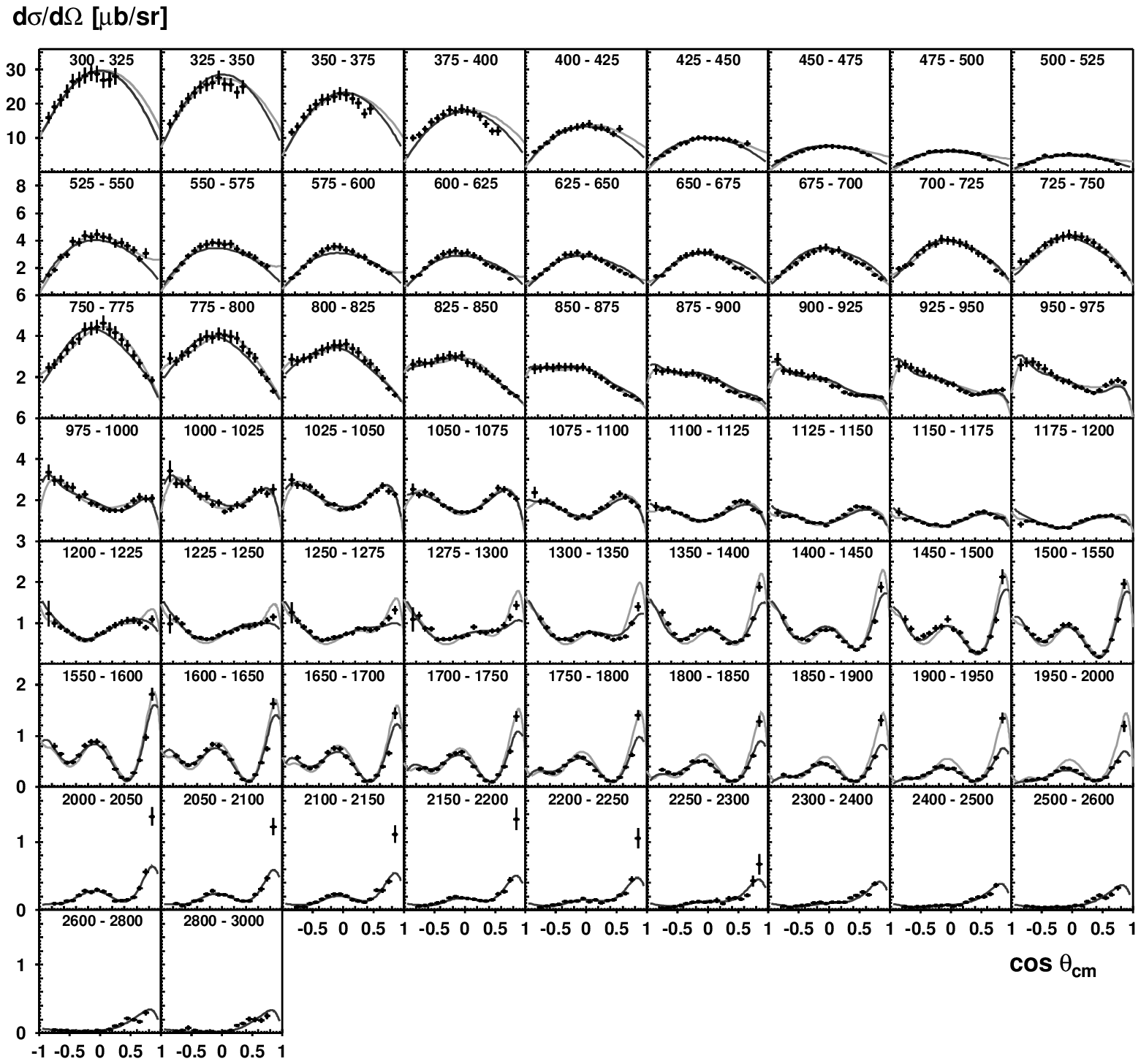}
\end{minipage}
\hspace*{3.5cm}
\begin{minipage}{11.5cm}
\vspace{-32mm}
\caption{\label{figure_pi0dcs}Differential cross sections for
$\rm\gamma p\rightarrow p\pi^0$ ($E_{\gamma}$ bins in MeV).  
\protect\phantom{The measured}
The measured 
data points are shown 
as black squares, the PWA result as a dark, the SAID parameterization 
as light gray line. At large $E_{\gamma}$, SAID reproduces the very forward 
data points, yet the agreement over the full angular range is worse.}
\end{minipage}
\vspace{-12mm}
\end{figure*}
\par
The absolute normalization was derived from the measured photon flux
and by a comparison of our data with SAID.  For each
energy channel in the tagger, the angular distributions were divided
into 20 bins in $\cos\theta_{\rm cm}$
where $\theta_{\rm cm}$ denotes
the emission angle
of the meson in the center--of--mass system with respect to the initial
photon direction.  For the low
energy data (below 1.3\,GeV) the resulting distributions were fitted to
SAID by the $\chi^2$ method for each energy channel.  In general, 
the shapes of the distributions are in excellent agreement.
The absolute flux normalization was determined from the SAID data base.
For the high--energy
data, this method could have been applied up to about 2\,GeV.
Above, experimental and
SAID angular distributions deviate significantly.
Below 2\,GeV, a comparison of our angular distributions and SAID defines
the absolute normalization of our photon flux, with one scaling factor
over the full energy range.
The uncertainty of the normalization is estimated to 5\%
for photon energies below 1.3\,GeV and to 15\% above.
\par
\begin{figure}
\vspace*{-0.3cm}
\epsfig{file=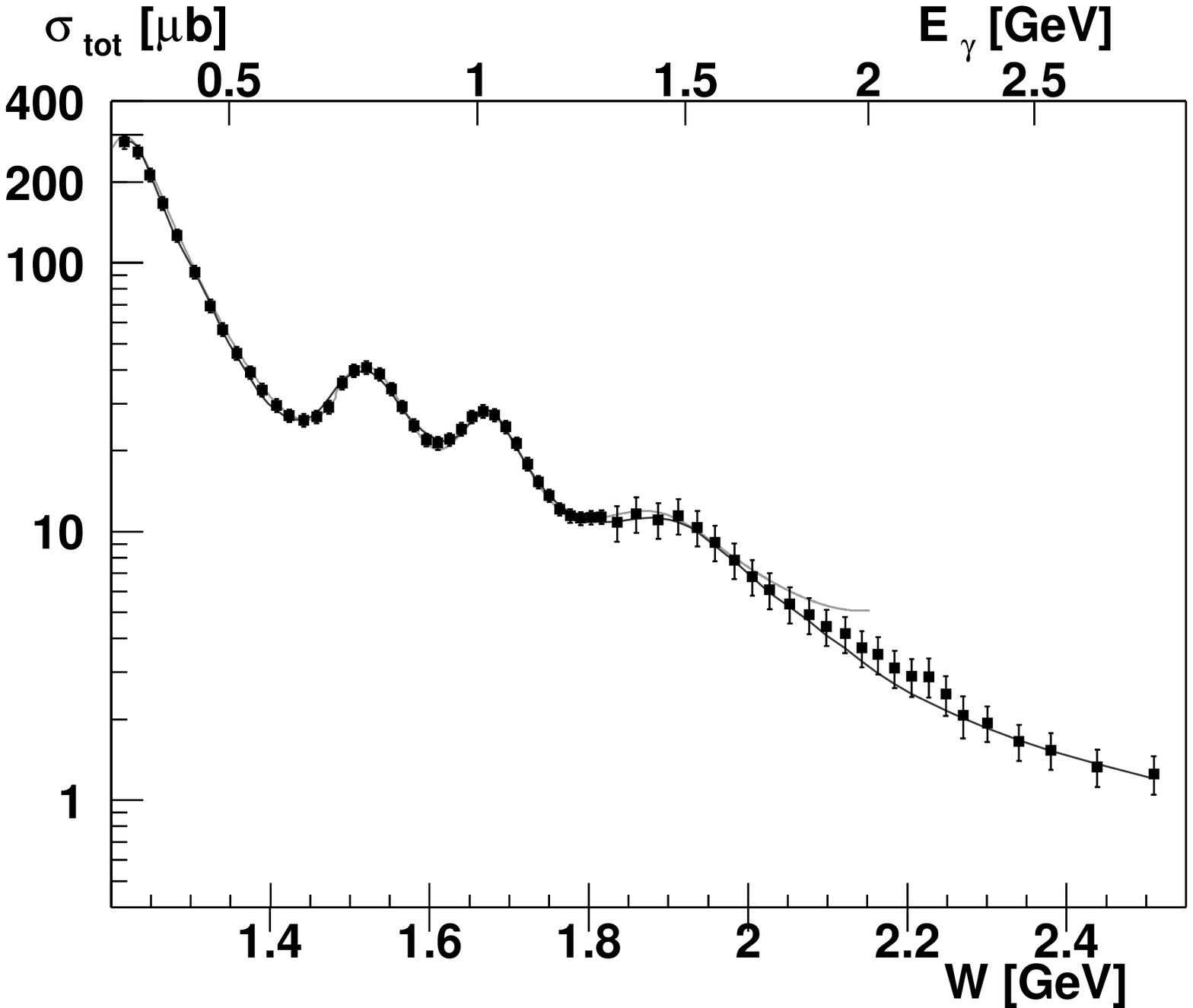, width=7.5cm}
\vspace*{-0.4cm}
\caption{\label{figure_pi0tcs} Total cross section for $\rm\gamma
p\rightarrow p\pi^0$ obtained by integration of angular distributions
and extrapolation into forward and backward regions using of PWA
result. The dark gray line represents the result of the PWA, the light 
gray line represents SAID.}
\vspace*{-5mm}
\end{figure}

The differential cross sections d$\sigma$/d$\Omega$ are shown in
Fig.~\ref{figure_pi0dcs}.  Statistical and
systematic errors are added quadratically.
The latter were calculated from Monte Carlo simulations of
different experimental uncertainties, a
shift of the photon beam by 3\,mm with respect to
the target center, a target shift of $\pm 1.5$\,mm in beam direction
and a change in the thickness of material between barrel and inner
detector corresponding to 1\,mm of capton foil.
The relative reconstruction efficiency error was estimated to 5\%.
As the angular
distributions are not influenced by the flux errors, these are not
included in Fig.~\ref{figure_pi0dcs}.
The dark gray line shows the result of
the partial wave analysis described below.
The total cross section in Fig.~\ref{figure_pi0tcs} was obtained by
integration of the measured cross section and extrapolation to the
non--measured region by using the results of the partial wave
analysis (PWA).
\par
The data was subjected to a coupled--channel partial wave analysis
within the framework of an isobar model.  Resonances are described by
Breit--Wigner amplitudes, except for strongly overlapping states which
are described in K--matrix formalism. Angular distributions are
calculated in the spin--orbital decomposition
formalism~\cite{Rarita:mf,Anisovich:tu}. The
PWA method is described in~\cite{AST}.
\par
Along with the CB--ELSA data on photoproduction of $\pi^0$ and $\eta$
mesons as discussed here and in the subsequent
letter~\cite{Crede:2003ax},  several other data sets served as input
for the analysis:  The TAPS low--energy data on
$\rm\gamma p\rightarrow p\eta$~\cite{Krusche:1995} to fix the $\eta$ threshold;
beam--asymmetry measurements of $\pi^0$~\cite{GRAAL1,SAID1}
and $\eta$~\cite{GRAAL1,GRAAL2}
photoproduction were used to provide
additional constraints from polarization observables. The beam--asymmetry
data covers the mass range from 1200 to 2000\,MeV.  We
include data on $\rm\gamma p\to n\pi^+$~\cite{SAID2} covering the mass
range from 1500 to 2150\,MeV thus exploiting the different decay
couplings of N$^*$ and $\Delta^*$ resonances to n$\pi^+$ and
p$\pi^0$.
\par
The $\chi^2$ of the final fit values for the different data sets are given in~\cite{Crede:2003ax}.
The masses, widths, helicity ratios, and contributions of different resonances
to the cross section for $\pi^0$ photoproduction from the final solution are
given in Tab.~\ref{resonance_list}.  We consider contributions to the total cross section with
fractional yields (column 5) above 1--2\%
as established from this analysis.  The sum of fractional yields is not equal to one due to interferences.
\begin{table}[h]
\caption{Masses, widths and resonance couplings, this analysis.}
\renewcommand{\arraystretch}{1.1}
\begin{tabular}{lllcc}
\hline\hline
Resonance& {\footnotesize{ M (MeV)}}   & {\footnotesize{$\Gamma$ (MeV)}}&
 {\footnotesize{$A_{1/2}/A_{3/2}$}} & Fraction \\
\hline
$\rm N(1440)P_{11}$&$1375\pm 40$&$380\pm 120$    & --              & 0.037 \\
$\rm N(1520)D_{13}$&$1523\pm 4$&$105^{+6}_{-18}$&$~~0.08\pm 0.10$& 0.108\\
$\rm N(1535)S_{11}$$^*$  &$1501\pm 5$&$215\pm 25$& \multirow{2}*{--} & \multirow{2}*{0.150}\\
$\rm N(1650)S_{11}$$^*$  &$1610\pm 10$&$190\pm 20$&               &  ~   \\
$\rm N(1675)D_{15}$&$1690\pm 12$&$125\pm 20$    &$~~0.06\pm 0.18$& 0.002 \\
$\rm N(1680)F_{15}$&$1669\pm 6 $&$ 85\pm 10$    &$-0.12\pm 0.04$ & 0.056 \\
$\rm N(1700)D_{13}$&$1740\pm 12$&$ 84\pm 16$    &$~~0.01\pm 0.20$& 0.001 \\
$\rm N(1720)P_{13}$&$1775\pm 18$&$325\pm 25$    &$~~0.68\pm 0.10$& 0.013 \\
$\rm N(2000)F_{15}$&$1950\pm 25$&$230\pm 45$    &$~~1.08\pm 0.60$& 0.002 \\
$\rm N(2070)D_{15}$&$2068\pm 22$&$295\pm 40$    &$~~1.37\pm 0.24$& 0.008 \\
$\rm N(2080)D_{13}$&$1943\pm 17$&$ 82\pm 20$    &$~~0.97\pm 0.28$& 0.000 \\

$\rm N(2200)P_{13}$&$2214\pm 28$&$360\pm 55$    &$~~0.41\pm 0.22$& 0.000  \\
\hline
$\rm\Delta(1232)P_{33}$$^{\diamond}$&$1235\pm 6 $&$135\pm 15$    &$~~0.52\pm 0.03$& 0.820 \\
$\rm\Delta(1620)S_{31}$&$1627\pm 8 $&$118\pm 15$& --              & 0.014\\
$\rm\Delta(1700)D_{33}$&$1686\pm  8$&$188\pm 18$    &$~~0.94\pm 0.12$& 0.077 \\
$\rm\Delta(1905)F_{35}$&$1970\pm 40$&$325\pm 50$    &$~~1.32\pm 0.45$& 0.004 \\
$\rm\Delta(1920)P_{33}$&$2032\pm 16$&$395\pm 50$    &$~~0.35\pm 0.12$& 0.032 \\
$\rm\Delta(1940)D_{33}$&$1910^{+20}_{-40}$
                           &$230\pm 45$    &$~~1.20^{+0.8}_{-0.4}$& 0.004 \\
$\rm\Delta(1950)F_{37}$&$1888\pm 6 $&$220\pm 18$    &$~~0.80\pm 0.11$& 0.037 \\
\hline\hline
\end{tabular}\\
\footnotesize{$^*$ K matrix fit, pole position
of the scattering amplitude in the\\~~complex plane, fraction for the total K--matrix contribution}\\
\footnotesize{$^{\diamond}$ This contribution includes non--resonant background.~~~~~~~}
\renewcommand{\arraystretch}{1.0}
\label{resonance_list}
\end{table}
\par
\par
In the total cross section clear peaks are observed for the first,
second, and third resonance region. A further broad enhancement can be
seen at about 1900\,MeV. The first resonance region is of course
mainly due to $\rm\Delta(1232)P_{33}$; the $\rm N(1440)P_{11}$
Roper resonance provides a small contribution of about 3--4\% compared to 
$\Delta(1232){\rm P}_{33}$. The $A_{1/2}/A_{3/2}$ helicity ratio
of excitation of $\rm\Delta(1232)P_{33}$ is determined to be 
$0.52\pm 0.03$ which agrees
favorably with the PDG average $0.53\pm 0.03$.
\par
The second resonance region receives approximately equal contributions
from $\rm N(1535)S_{11}$ and $\rm N(1520)D_{13}$. The helicity
ratio $|A_{1/2}/A_{3/2}|$ of the excitation of the $\rm N(1520)D_{13}$ 
is determined to be
small ($\leq 0.18$). From the couplings given by the PDG
we calculate $-(0.14\pm 0.06)$; Mukhopadhyay used an effective Lagrangian
approach to determine it to be $-(0.4\pm 0.1\pm 0.1)$~\cite{mukho}.
Within $\sim 2\sigma$ there is consistency between these values.
\par
The third bump in the data is made up out of three major
contributions. The $\rm\Delta(1700)D_{33}$ provides the largest fraction
($\sim\! 35$\%) of the peak, followed by $\rm N(1680)F_{15}$  ($\sim\!
25$\%) and $\rm N(1650)S_{11}$ ($\sim\! 20$\%)
as extracted from K--matrix parameterization;
$\rm\Delta(1620)S_{31}$ ($\sim\! 7$\%)
and $\rm N(1720)P_{13}$  ($\sim\! 6$\%) are also observed.
In the fourth resonance region we identify 
$\rm\Delta(1950)F_{37}$ contributing $\sim\! 41$\%
to the enhancement and $\rm\Delta(1920)P_{33}$ with
$\sim\! 35$\%.
\par
The fit also suggests the presence of
$\rm\Delta(1905)F_{35}$ and  $\rm\Delta(1940)D_{33}$.
Additionally, weak evidence is found for the new resonance
$\rm N(2070)D_{15}$ reported in \cite{Crede:2003ax} in the 
$\rm p\pi^0$ decay mode ($\sim\! 9$\%).  Here we mention only
that $\rm N(2070)D_{15}$ is predicted by quark models as being
one of the missing states. In \cite{Capstick:1993kb}, a doublet
of $\rm ND_{15}$ resonances is predicted at 2080 and 2095\,MeV, respectively,
where the lower (and only the lower) mass state has a large coupling to
N$\eta$. A similar mass pattern is found in \cite{Loring:2001kx}. 
The high--energy region is dominated by $\rho$--$\omega$ exchange in the
$t$ channel as can be seen by the forward peaking in the
differential cross sections.
\par
We have reported new data on the photoproduction of neutral pions off
protons over the full baryon resonance region. The data
extends the existing data base in larger solid angle coverage
and in energy. A partial wave analysis
of the data presented here and elsewhere~\cite{Crede:2003ax,AST}
decomposes the data into contributions from a large number
of N$^*$ and $\Delta^*$ resonances.
\begin{acknowledgments}
We thank
the technical staff at ELSA
and at all the participating institutions for their invaluable
contributions to the success of the experiment. We acknowledge
financial support from the Deutsche Forschungsgemeinschaft (DFG).  The
collaboration with St. Petersburg received funds from DFG and the
Russian Foundation for Basic Research.
\mbox{B.~Krusche} acknowledges
support from Schweizerischer Nationalfond. U.~Thoma thanks for an Emmy
Noether grant from the DFG. A.\,V.~Anisovich and A.\,V.~Sarantsev
acknowledge support from the Alexander von Humboldt Foundation. This
work comprises part of the PhD theses of O.~Bartholomy and H.~van~Pee.
\end{acknowledgments}

\end{document}